\documentclass{article}
\pdfoutput=1

\usepackage{arxiv}
\usepackage{amsmath}

\usepackage{algorithm}
\usepackage{algcompatible}
\usepackage{adjustbox}
\usepackage{amsfonts}
\usepackage{amssymb}
\usepackage{amsbsy}
\usepackage[T1]{fontenc}
\usepackage{textcomp}
\usepackage{graphicx}
\usepackage{listings}
\usepackage{multirow}
\usepackage[normalem]{ulem}
\usepackage{color}
\definecolor{ruby}{rgb}{0.6,0,0.3}
\definecolor{gold1}{rgb}{0.8667,0.8510,0.7647}
\definecolor{gold2}{rgb}{0.7686,0.7412,0.5922}
\definecolor{gold3}{rgb}{0.5804,0.5412,0.3294}
\definecolor{blue1}{rgb}{0.7765,0.8510,0.9451}
\definecolor{blue2}{rgb}{0.5569,0.7059,0.8902}
\definecolor{blue3}{rgb}{0.3333,0.5569,0.8353}
\definecolor{gray1}{rgb}{0.9,0.9,0.9}
\definecolor{gray2}{rgb}{0.8,0.8,0.8}
\definecolor{gray3}{rgb}{0.7,0.7,0.7}

\usepackage{soul}
\newcommand{\hlc}[2][gold1]{ {\sethlcolor{#1} \hl{#2}} }

\usepackage{epstopdf}
\usepackage[caption=false]{subfig}

\usepackage{multirow}
\usepackage{xparse}
\ExplSyntaxOn
\NewDocumentCommand{\smallcaps}{m}
 {
  \tl_set:Nn \l_tmpa_tl { #1 }
  \regex_replace_all:nnN
   { ([0-9]+) } 
   { \c{resizedigit}\cB\{ \1 \cE\} } 
   \l_tmpa_tl
  \textsc{ \tl_use:N \l_tmpa_tl }
 }

\ExplSyntaxOff

\usepackage{tikz}
\newcommand*\circled[1]{\tikz[baseline=(char.base)]{
            \footnotesize\node[shape=circle,draw,inner sep=1pt] (char) {#1};}}

\usepackage[hidelinks]{hyperref}
\usepackage{url}

\graphicspath{{../figures/}}
\newcommand{\ignore}[1]{}
\newcommand{\revisit}[1]{}

\title{GPU implementation of a ray-surface intersection algorithm in CUDA (Compute Unified Device Architecture)}
\shorttitle{GPU implementation of a ray-surface intersection algorithm in CUDA}

\date{September 6, 2022}	

\author{
  \textbf{Raymond~Leung}\vspace{2mm} \\
  Australian Centre for Field Robotics\\
  Faculty of Engineering\\
  The University of Sydney\\ \vspace{-10mm}
  \ignore{\texttt{raymond.leung@sydney.edu.au} \\}
}

\renewcommand{\headeright}{}
\renewcommand{\undertitle}{}

\begin{document}
\maketitle

\begin{abstract}
These notes accompany the open-source code published in \href{https://github.com/raymondleung8/gpu-ray-surface-intersection-in-cuda}{\color{blue}GitHub} which implements a GPU-based line-segment, surface-triangle intersection algorithm in CUDA. It mentions some relevant works and discusses issues specific to this implementation. The goal is to provide software documentation and greater clarity on collision buffer management which is sometimes omitted in online literature. For real-world applications, CPU-based implementations of the test are often deemed too slow to be useful. In contrast, the code described here targets Nvidia GPU devices and offers a solution that is vastly more efficient and scalable. The main API is also wrapped in Python. This geometry test is applied in various engineering problems, so the software developed can be reused in new situations.\footnote{The practical applications of the ray-surface intersection test are well understood in computer graphics. For instance, it is used to render photo-realistic objects in a scene \cite{pharr2016physically} and evaluate user cursor interactions with a 3D terrain height map \cite{sjostrand2017b}. Apart from these examples, it is also used in spatial algorithms in geoscience, for instance, to label geological domains in orebody grade-block models \cite{leung2020mos}. Its reach extends to fabrication of nanoelectronic devices and 3D topography simulation for etching and deposition processes, where ray-surface intersection tests are used to model particle transport and surface evolution \cite{manstetten2018phd}.}
\end{abstract}

\section{Motivation}
The objective is to determine for each line segment $l_i=(\mathbf{r}_{i}^\text{start},\mathbf{r}_{i}^\text{end})$ whether it crosses a user-supplied surface, given $N_\text{r}$ rays, each described by start and end points $\mathbf{r}_{i}^\text{start}\in\mathbb{R}^3$ and $\mathbf{r}_{i}^\text{end}\in\mathbb{R}^3$, and a mesh surface that comprises $N_\text{t}$ triangles, each described by a triplet $\mathbf{t}_j=[t_{j,1},t_{j,2},t_{j,3}]$ which references the three vertices $\mathbf{v}(\mathbf{t}_j)\equiv[\mathbf{v}_{t_{j,1}},\mathbf{v}_{t_{j,2}},\mathbf{v}_{t_{j,3}}]$ from the collection $\mathcal{V}=\{\mathbf{v}_n\}_{1\le n\le N_\text{v}}$. The result is stored in a boolean array, $\mathbf{y}\in\{0,1\}^{N_\text{r}}$ by default. However, the program can be configured to return the intersecting point and triangle for each surface-intersecting ray. In the envisaged application, ray segments can point in arbitrary directions. Their starting points $\mathbf{r}_{i}^\text{start}$ may be all different. There is no requirement for the rays to originate from a single or multiple shared light sources.

The principal motivation for writing the ray-surface intersection test code in C++/CUDA is to harness the power of GPUs \cite{brodtkorb2013graphics} and increase productivity in an R\&D project that is geared towards geotechnical investigations\,/\,stratigraphic modelling. As such, achieving real-time rendering performance (e.g. 25 fps) is not the real goal, rather the goal is to surpass the speed of off-the-shelf CPU-based solutions, such as python's {\small\texttt{vtk}} and {\small\texttt{pyvista}} packages. For instance, the latter takes $\sim 60$s to test 10M rays against a surface with approx. 30k triangles, even with OpenMP under the hood.\footnote{OpenMP (Open Multi-Processing) is an application programming interface (API) that supports multi-platform shared-memory multiprocessing programming in C, C++, and Fortran, on many platforms, instruction-set architectures and operating systems, including Solaris, AIX, FreeBSD, HP-UX, Linux, macOS, and Windows. It consists of a set of compiler directives, library routines, and environment variables that influence run-time behavior (Wikipedia, 2022) \cite{openmp-wiki}.} A reasonable goal is to reduce this duration by 10 to 100 fold. In general, GPU programming requires a different mindset and conceptual organisation; a good summary of its philosophy and applications can be found in \cite{owens2008gpu}.

\section{Implementation}
Our initial CUDA design choices were informed by findings reported by Jimenez et al. in \cite{jimenez2014performance} which influenced how the problem was partitioned and translated to CUDA architecture \cite{de2010introduction,kirk2016programming}. In accordance, each grid ($j$) loads data pertaining to a single triangle ($\mathbf{t}_j$), the first thread in each thread-block computes shared attributes such as the edges of the triangle with a synchronisation barrier. The thread index ($i$) is mapped to a ray-segment ($l_{i+\lambda B}$), it cycles over multiplier $\lambda\in\mathbb{Z}$ until all rays have been considered, with stride $B=1024$ being equal to the block size. Each thread is responsible for applying the Moller-Trumbore test to each triangle-ray ($j, i+\lambda B$) combination. A number of algorithms---Badouel, Moller-Trumbore, Segura-Feito (tetrahedra sign test) and Jimenez et al. (barycentric coordinates of segment endpoint)---were considered in \cite{jimenez2010robust} and the study in \cite{jimenez2014performance} demonstrated that Moller-Trumbore \cite{moller1997fast} is the most efficient general-purpose algorithm for detecting ray-triangle intersections, requiring only 1 division, 4 dot-products and 2 vector-products in the worst case. Although not specifically mentioned by Jimenez, our initial implementation also included a na\"ive \textit{screening or culling} step that checks for ray-triangle AABB (axis-aligned bounding box) overlap before the exact test is applied. A major conclusion in Jimenez's study \cite{jimenez2014performance} was that precalculating quantities is ``typically counterproductive in the GPU environment'' as the time required to compute these values may be smaller than the time required to fetch them from memory. In our experience, if the ray-segment bounding boxes were precomputed in the context of the bounding-box prescreening strategy, it increases the throughput roughly by 25\%. In this preliminary implementation (v1), the cost is still significant even with the screening step, as it applies the overlap test to all ($N_\text{t}\times N_\text{r}$) combinations. Nonetheless, it is faster than some of CPU-based alternatives. This is shown in Table~\ref{tab:compare1} where speed is measured in million intersection tests per second (Mi/s) using an Nvidia GTX Titan X GPU.

\begin{table}[h!]
\begin{center}
\renewcommand{\arraystretch}{1.1}
\caption{Timing comparison for the preliminary implementation}\label{tab:compare1}
\begin{tabular}{|l|c|c|c|}\hline
& Time (s) & Speed (Mi/s) & GPU gain\\ \hline
\multicolumn{4}{|l|}{\textbf{CPU-based solutions}}\\ \hline
PyVista ({\small\texttt{ray\_trace}}) & 384.5 & 761.5 & $16.2\times$\\
PyVista ({\small\texttt{multi\_ray\_trace}}) OpenMP with 8 logical processors & 60.6 & 4832.5 & $2.55\times$\\
VTK ({\small\texttt{obbTree.IntersectWithLine}}) & 92.3 & 3171.3 & $3.89\times$\\ \hline
\multicolumn{4}{|l|}{\textbf{Our preliminary GPU CUDA implementation}}\\ \hline
Jimenez-like strategy (v1) & 23.7 & \textbf{\color{ruby}12340.5} &\\ \hline
\multicolumn{4}{|l|}{\textbf{Reference GPU implementation}}\\ \hline
{\color{gray}Jimenez \cite{jimenez2014performance} Table 2 (original reported figure)} & -- & {\color{gray}2742.0}$^\dag$ &\\
Jimenez \cite{jimenez2014performance} Table 2 (hardware adjusted) & -- & \textbf{9651.9}$^\ddag$ & $1.27\times$\\ \hline
\multicolumn{4}{l}{$^\dag$ Reported figure using GeForce GTX560Ti.}\\
\multicolumn{4}{l}{$^\ddag$ Adjusted figure using an \href{https://gpu.userbenchmark.com/Compare/Nvidia-GTX-Titan-X-vs-Nvidia-GTX-560-Ti/3282vs2180}{estimated speedup factor of 3.52} assuming GTX Titan X is used today.}\\
\multicolumn{4}{l}{Test surface contains 29,284 triangles, $1\times10^7$ (10 million) rays were used.}\\
\end{tabular}
\end{center}
\end{table}

All this illustrates is that our preliminary CUDA implementation \cite{leung2022gpursi} and Jimenez' reference implementation achieve quite similar performance. To achieve more significant speedup, the screening step cannot be applied in a brute force manner to all possible ray-triangle pairs. Embedding location information about the triangles in a hierarchical structure is key to avoiding needless evaluations. Hence, our final implementation (version v2) makes use of more advanced data structures and performs tree search to eliminate possibilities of overlap. Online resources provide excellent practical guidance. Our approach is guided by Marcus \cite{marcus2015} which in turn is based on Apetrei's fast, agglomerative approach \cite{apetrei2014fast} for building a linear bounding volume hierarchy (LBVH \cite{bvh-wiki}). The idea is to construct a binary radix tree for a set of objects (viz., triangles on the mesh surface) which are sorted using morton codes. This allows the tree structure to be traversed in parallel by individual threads, each associated with a particular ray-segment. A common choice is to use the centroid of the triangle vertices, $\mathbf{c}_j=[\mathbf{v}_{t_{j,1}}\!+\!\mathbf{v}_{t_{j,2}}\!+\!\mathbf{v}_{t_{j,3}}]/3$, to represent the location of each triangle.

The first step is to compute the support intervals over $\{\mathbf{c}_j\}$. Then, each triangle centroid, $\mathbf{c}_j\in\mathbb{R}^3$, is quantised as $\mathbf{q}_j\in\mathbb{Z}^3$ to utilise the range effectively (with up to $2^{21}$ levels along the x, y and z axes). The quantised triangle coordinates, $[q_{j,x}, q_{j,y}, q_{j,z}]$, are subsequently bit-wise interleaved and converted into 64-bit long unsigned Morton code, $\mathbf{z}_j$. This has the effect that when all the codes ($\mathbf{z}_j$) are sorted, spatial locality will be preserved, which essentially means the triangles are grouped\,/\,clustered locally following a Z-curve scanning pattern \cite{mortoncode-wiki} in 3D space. Once the triangles $\mathbf{q}_t\equiv Q(\mathbf{c}_t)$ are mapped to Morton codes $\mathbf{z}_j$, they are sorted to correspond to leaf nodes. The second step is to construct the binary radix tree (LBVH) by setting up left and right child node pointers for each parent in a bottom-up manner starting from the leaf nodes (each associated with an individual triangle). Apetrei \cite{apetrei2014fast} devised a single-pass procedure for propagating the index range and bounding box values from children to parents. This construction is implemented as GPU device code; for details, refer to the {\small\texttt{bvhConstruct}} kernel in ``{\small\texttt{bvh\_structure.h}}''.

Once the LBVH is built, there needs to be a way for querying bounding-box overlap with individual line-segment bounding box using the triangle binary radix tree. The relevant structures for the latter are stored in two device memory arrays: {\small\texttt{internalNodes}} and {\small\texttt{leafNodes}}. On paper, the size of these are always $N_\text{t}-1$ and $N_\text{t}$. However, it is convenient to actually store the \textit{root node} as the left child of the last internal node (in {\small\texttt{internalNodes[$N_\text{t}\!-\!1$]}}). This pointer is accessed as a shared variable in the relevent kernel launch ({\small\texttt{bvhIntersectionKernel}}) which provides the starting point for searching the tree. At the thread-block level, each thread executes the device code {\small\texttt{bvhFindCollisions(..., idx)}}. This GPU code plays two major roles: a) it conducts efficient overlap feasibility testing using {\small\texttt{bvhTraverse}} (iterative tree search) to identify a small set of triangles capable of intersecting with each ray; b) it invokes the {\small\texttt{lib\_rsi::checkRayTriangleIntersection}} function to verify if there is an actual crossing between the ray and surface. Pictures and further explanation of the BVH traversal function can be found in (Kerras, 2012) \cite{karras2012}.

\subsection{Program structure}
The code repository contains four essential files.
\begin{itemize}
\item \textbf{gpu\_ray\_surface\_intersect.cu} contains the {\small\texttt{main}} program which handles I/O, allocates host/device memory and launches the relevant kernels.
\item \textbf{bvh\_structure.h} implements the bounding volume hierachy in the {\small\texttt{lib\_bvh}} namespace. It defines \_\_device\_\_ functions and provides \_\_global\_\_ functions (such as bvhConstruct and bvhIntersectionKernel) used in {\small\texttt{main}}.
\item \textbf{rsi\_geometry.h} implements the Moller-Trumbore ray-triangle intersection algorithm (and associated algebraic operations) in the {\small\texttt{lib\_rsi}} namespace. It provides the {\small\texttt{checkRayTriangleIntersection}} interface function used in {\small\texttt{lib\_bvh::bvhFindCollisions}}.
\item \textbf{morton3D.h} inherits some of the Morton code procedures written by Jeroen Baert which is distributed under an MIT license.
\end{itemize}

\subsection{Compilation}
The code is compiled using\\
{\small\texttt{{\color{gray}/usr/local/cuda/bin/}nvcc gpu\_ray\_surface\_intersect.cu -o gpu\_ray\_surface\_intersect}}

\subsection{Standard usage}
The basic command is\\
{\small\texttt{./gpu\_ray\_surface\_intersect \$\{vertices\_file\} \$\{triangles\_file\} \$\{rayfrom\_file\} \$\{rayto\_file\}}}

\noindent If the input files are named ``vertices\_f32'', ``triangles\_i32'', ``rayFrom\_f32'', ``rayTo\_f32'' and located in {\small\texttt{./input}} relative to the source directory, the last four command line arguments may be omitted.

To \textbf{suppress terminal output}, the string {\small\texttt{silent}} may be added as the 5\textsuperscript{th} argument.\\
{\small\texttt{./gpu\_ray\_surface\_intersect \$\{vertices\} \$\{triangles\} \$\{rayfrom\} \$\{rayto\} {\color{ruby}silent}}}

\subsection{Extended usage}\label{sec:extended-usage}
To \textbf{return results in the form of\hlc[gray2]{(intersecting\_rays, distances, intersecting\_triangles, intersecting\_points)}} in lieu of a {\small\texttt{crossing\_detected}} binary array, the string {\small\texttt{barycentric}} is added as the 6\textsuperscript{th} argument.\\
{\small\texttt{./gpu\_ray\_surface\_intersect \$\{vertices\} \$\{triangles\} \$\{rayfrom\} \$\{rayto\} default {\color{ruby}barycentric}}}

To \textbf{return results in the form of\hlc[gray2]{(num\_ray\_surface\_intersecting\_points)}}, use\\
{\small\texttt{./gpu\_ray\_surface\_intersect \$\{vertices\} \$\{triangles\} \$\{rayfrom\} \$\{rayto\} default {\color{ruby}intercept\_count}}}

\subsection{Required input}
The input consists of four binary files in little-endian format.
\begin{itemize}
\item \textbf{vertices} contains $3\times N_\text{vertices}\times \text{sizeof}({\small\texttt{float32}})$ bytes. The $N_\text{vertices}$ surface vertices are arranged as $v_{1,x},v_{1,y},v_{1,z},v_{2,x},v_{2,y},v_{2,z},...$.
\item \textbf{triangles} contains $3\times N_\text{triangles}\times \text{sizeof}({\small\texttt{int32}})$ bytes. The $N_\text{triangles}$ surface triangles are arranged as $t_{1,1},t_{1,2},t_{1,3},t_{2,1},t_{2,2},t_{2,3},...$.
\item \textbf{rayFrom} contains $3\times N_\text{rays}\times \text{sizeof}({\small\texttt{float32}})$ bytes. The $N_\text{rays}$ line-segment start points are arranged as $r_{1,x}^\text{start},r_{1,y}^\text{start},r_{1,z}^\text{start},r_{2,x}^\text{start},r_{2,y}^\text{start},r_{2,z}^\text{start},...$.
\item \textbf{rayTo} contains $3\times N_\text{rays}\times \text{sizeof}({\small\texttt{float32}})$ bytes. The $N_\text{rays}$ line-segment end points are arranged as $r_{1,x}^\text{end},r_{1,y}^\text{end},r_{1,z}^\text{end},r_{2,x}^\text{end},r_{2,y}^\text{end},r_{2,z}^\text{end},...$.
\end{itemize}
Sample input data may be generated using {\small\texttt{scripts/input\_synthesis.py}} \cite{leung2022gpursi}.

\subsection{Python wrapper}
For convenience, \textbf{scripts/gpu\_ray\_surface\_intersect.py} implements a wrapper class {\small\texttt{PyGpuRSI}} that encapsulates the functionality of {\small\texttt{gpu\_ray\_surface\_intersect.cu}}. The notebook {\color{ruby}\small\texttt{scripts/demo.ipynb}} illustrates how this is used. Part A shows how the data manipulation, compilation, run and clean-up steps can be managed using a {\small\texttt{with}} statement. Part B shows how the program can be configured to return {\small\texttt{(intersecting\_rays, distances, intersecting\_triangles, intersecting\_points)}} as python objects (numpy arrays) in reference to the extended usage described in Sec.~\ref{sec:extended-usage}. The notebook {\color{ruby}\small\texttt{scripts/experimental\_feature.ipynb}} contains an example in Part C where {\small\texttt{(num\_intersecting\_points)}} is returned instead. The odd parity test may be used to find ray starting points that lie inside a closed surface.

\section{Discussion}
In this section, we comment on specific aspects of the implementation that may seem a little obscure.

\subsection{Bounding box and node definitions}
The axis-aligned bounding box (AABB) and node are defined in {\small\texttt{rsi\_geometry.h}} and {\small\texttt{bvh\_structure.h}}, respectively.

\begin{lstlisting}[frame=single,emph={},emphstyle=\textbf,basicstyle={\ttfamily\footnotesize},escapechar=@]
struct @\textbf{AABB}@
{
    float xMin, xMax, yMin, yMax, zMin, zMax;
};

struct @\textbf{BVHNode}@
{
    AABB bounds;
    BVHNode *childLeft, *childRight;
    BVHNode *parent; @$\star$@
    BVHNode *self;   @$\star$@
    int idxSelf, idxChildL, idxChildR, isLeafChildL, isLeafChildR; @$\star$@
    int triangleID;
    int atomic;
    int rangeLeft, rangeRight;
};
\end{lstlisting}

Most of these fields are explained in Robbin Marcus' blog \cite{marcus2015}, thus there is no point repeating except in noting that attributes marked with $\star$ are not strictly necessary. These are included for debugging purpose during development to verify that {\small\texttt{parent}} points correctly to the parent node and one of its children nodes points to the current node {\small\texttt{self}}, for example, when we trace and expand a portion of the tree using the {\small\texttt{testSimulateTreeExpansion}} function and examines its contents with {\small\texttt{testPrintNode}}. The reason for having the redundant variables {\small\texttt{idxSelf, idxChildL, idxChildR, isLeafChildL, isLeafChildR}} is to explicitly identify the array indices and knowing whether {\small\texttt{idxChild?}} refers to an element in the {\small\texttt{internalNodes}} or {\small\texttt{leafNodes}}. This is because the {\small\texttt{BVHNode}} pointers become invalid when the structure is copied from device to host memory, dereferencing them would lead to illegal memory access, so instead we use {\small\texttt{testDecipherDescendent}} to interpret {\small\texttt{idxChild?}} to find the corresponding address in host memory.

The attribute {\small\texttt{bounds}} obviously describes the spatial extent of the bounding box for {\small\texttt{triangles[triangleID]}} while {\small\texttt{[rangeLeft, rangeRight]}} describes the range covering the node indices of its descendants. In our implementation, an internal node is assigned a {\small\texttt{triangleID}} value of -1, the root node in particular is given a unique value of -2. All leaf nodes have non-negative values $0 \le {\small\texttt{triangleID}} < N_\text{t}$. The {\small\texttt{atomic}} variable is used during BVH construction (see {\small\texttt{bvhUpdateParent}}) to ensure the internal nodes only compute the {\small\texttt{bounds}} from their children bounding boxes when both children have been identified.

\subsection{Bounding box collision detection}
The non-trivial part of the implementation revolves around {\small\texttt{bvhFindCollisions}} and {\small\texttt{bvhInsert}}. Although there is an excellent description of stack pointers in \cite{karras2012} which also explains the thinking behind BVH tree traversal in GPU, details about the management of the collision list are somewhat lacking. Our intention here is to highlight one feasible approach and discuss this in practical terms. Conceptually, the \textsc{traverse} step begins by comparing the bounding box of a given ray-segment (henceforth denoted $\Xi$) with the {\small\texttt{bounds}} in the current node $N_i$. The process starts at the root node, and when an overlap occurs, the left and right child nodes are expanded. If $\Xi$ intersects with the left child's bounding box, $\Omega_{L_i}$, and the left child happens to be a leaf node, {\small\texttt{node[$L_i$].triangleID}} is inserted into the collision buffer. Similarly, if $\Xi$ intersects with the right child's bounding box, $\Omega_{R_i}$ which happens to be a leaf node, {\small\texttt{node[$R_i$].triangleID}} is inserted into the collision buffer. This represents part 1 of 2 functions performed by {\small\texttt{bvhTraverse}}---refer to code between \circled{1}$\rightarrow$ and $\leftarrow$\circled{1}.

\begin{lstlisting}[frame=single,emph={},emphstyle=\textbf,basicstyle={\ttfamily\footnotesize},escapechar=@]
__device__ void @\textbf{\color{ruby}bvhTraverse}@(const AABB& queryBox, NodePtr &bvhNode,
                           NodePtr* &stackPtr, CollisionList &hits)
{
    NodePtr node(bvhNode);
    bool bufferFull(false);
    do @\circled{1}$\rightarrow$@
    {
        //check each child node for overlap
        NodePtr childL = node->childLeft;
        NodePtr childR = node->childRight;
        bool overlapL = overlap(queryBox, childL);
        bool overlapR = overlap(queryBox, childR);

        //query overlaps a leaf node => report collision
        if (overlapL && isLeaf(childL))
            bufferFull = bvhInsert(hits, childL->triangleID); @$\star$@

        if (overlapR && isLeaf(childR))
            bufferFull |= bvhInsert(hits, childR->triangleID); @$\leftarrow$\circled{1}@

        //query overlaps an internal node => traverse @\circled{2}$\rightarrow$@
        bool traverseL = (overlapL && !isLeaf(childL));
        bool traverseR = (overlapR && !isLeaf(childR));

        if (!traverseL && !traverseR)
            node = *--stackPtr; //pop
        else
        {
            node = (traverseL) ? childL : childR;
            if (traverseL && traverseR)
                *stackPtr++ = childR; //push
        } @$\leftarrow$\circled{2}@
    }
    while (node != NULL && !bufferFull);

    bvhNode = node;
}
\end{lstlisting}
The problem is that STL containers and adaptors (such as {\small\texttt{std::vector}} and {\small\texttt{std::stack|queue}}) cannot be used in device or kernel code as they lie outside the scope of a device function. Although it is possible to allocate memory dynamically and adjust the amount of device memory available for the heap using CUDA runtime APIs, this would be painfully slow if each thread allocates its own memory dynamically and memory is not coalesced. Thus, the collision buffer really needs to be statically allocated using {\small\texttt{cudaMalloc}}. Effectively, access to the relevant portion of the collision buffer is compartmentalised for each individual thread. The buffer size is finite and fixed. For instance, in the following listing, {\small\texttt{MAX\_COLLISION}} may be set to 32, say. Since there is no telling how many triangle bounding boxes would overlap with a given ray-segment, there needs to be a way to manage this {\small\texttt{triangleID}} \textsc{insert} operation to prevent array overrun. This is accomplished with {\small\texttt{bvhInsert}} which returns true when the {\small\texttt{hits}} buffer occupancy is nearing capacity. The ray-specific collision buffer occupancy state is reflected by the {\small\texttt{bufferFull}} thread-local variable in $\star$ above. The code between \circled{2}$\rightarrow$ and $\leftarrow$\circled{2} pops a node pointer from the {\small\texttt{stackPtr}} ``queue''\footnote{The stack pointer ``queue''  is actually a static array with known dimension at compile time.} when the ray bounding-box yields an empty intersection with \textbf{both} child bounding boxes. Otherwise, the first child node is visited next, and if the second child node also intersects with the ray-bounding box, this second child node is added to the {\small\texttt{stackPtr}} queue. Normally, this process continues until the ``queue'' is empty, i.e. when the NULL pointer is encountered. In our implementation, it also exits the do-while loop when a full buffer is imminent. Both input arguments {\small\texttt{NodePtr* \&bvhNode}} and {\small\texttt{CollisionList \&hits}} are mutable.

\begin{lstlisting}[frame=single,emph={},emphstyle=\textbf,basicstyle={\ttfamily\footnotesize},escapechar=@]
struct @\textbf{CollisionList}@
{
    uint32_t hits[MAX_COLLISIONS];
    int count;
};
__device__ bool inline @\textbf{\color{ruby}bvhInsert}@(CollisionList &collisions, int value)
{
    //insert value into the hits[] array. Returned value indicates
    //if buffer is full (true => not enough room for two elements).
    collisions.hits[collisions.count++] = static_cast<uint32_t>(value);
    return (collisions.count < MAX_COLLISIONS - 1)? false : true;
}
\end{lstlisting}

In the next listing, we see that {\small\texttt{bvhTraverse}} is wrapped in the device function {\small\texttt{bvhFindCollisions}} which performs two critical roles. First, it performs broad-based ray-triangle bounding box collision detection. Second, it performs the Moller-Trumbore check (an exact intersection test) on viable candidates (a small subset of triangles) for which a ray-triangle crossing is possible. The code in \circled{3} initialises the {\small\texttt{stackPtr}} array and sets up the {\small\texttt{bvhRoot}} as the first node to visit. This is consistent with the description in \cite{karras2012}. Where it differs, or perhaps just some design choices specific to our implementation, is the coupling with {\small\texttt{checkRayTriangleIntersection}} in \circled{4} instead of running coarse and fine-grained checks in separate stages. It applies the Moller-Trumbore test to the thread-specific ray-segment and a set of triangle candidates reported in the {\small\texttt{collisions}} list. If this list contains \textbf{all} potential collision candidates (the entire subset of triangles to test for), all is good. In the event the collisions buffer is full and there are more BVH nodes remaining to check, the {\small\texttt{nextNode}} returned will not be NULL; this means the do-while loop will continue unless a ray-triangle intersection has already been found by the Moller test. Another important observation is the internal state (or stack content) will persist in memory if the {\small\texttt{bvhTraverse}} function is called a second time within the do-while loop. As should be obvious, the ray-surface intersection result is affirmed in (written to) {\small\texttt{detected[rayIdx]}}.

\begin{lstlisting}[frame=single,emph={},emphstyle=\textbf,basicstyle={\ttfamily\footnotesize},escapechar=@]
__device__ void @\textbf{\color{ruby}bvhFindCollisions}@(const float* vertices,
                                  const int* triangles,
                                  const float* rayFrom,
                                  const float* rayTo,
                                  const AABB* rayBox,
                                  const NodePtr bvhRoot,
                                  CollisionList &collisions,
                                  int* detected,
                                  int rayIdx)
{   //argument `collisions` provides access to a thread-local
    //portion of device memory that holds the collisions array.

    //allocate traversal stack from thread-local memory,
    //push NULL to indicate that there are no postponed nodes.
    @\circled{3}$\rightarrow$@
    NodePtr stack[MAX_STACK_PTRS];
    NodePtr* stackPtr = stack;
    *stackPtr++ = NULL;
    NodePtr nextNode(bvhRoot);
    @$\leftarrow$\circled{3}@

    @\circled{4}$\rightarrow$@
    do {
        collisions.count = 0;
        bvhTraverse(rayBox[rayIdx], nextNode, stackPtr, collisions);

        //check for actual intersections with the triangles found so far
        int candidate = 0;
        while (! detected[rayIdx] && (candidate < collisions.count)) {
            int triangleID = collisions.hits[candidate++];
            checkRayTriangleIntersection(vertices, triangles, rayFrom, rayTo,
                                         detected, rayIdx, triangleID);
        }
    }
    while ((detected[rayIdx] == 0) && (nextNode != NULL));
    @$\leftarrow$\circled{4}@
}
\end{lstlisting}

To give an indication of the time cost associated with BVH construction in GPU, the following table shows the cumulative time as each step of the process is added. Overall, it is quite inexpensive for a test surface with about 29k triangles.
\begin{table}[h!]
\begin{center}
\renewcommand{\arraystretch}{1.25}
\caption{Timing associated with triangle BVH construction}\label{tab:bvh}
\begin{tabular}{|l|c|}\hline
Incremental steps & Elpased time (ms)\\ \hline
rbxKernel (compute ray bounding boxes) & 4.32\\
+ discretisation\,/\,normalisation of triangle coordinates & 5.47\\
+ create Morton code for triangles & 7.75\\
+ sort Morton code for triangles & 24.23\\
+ bvhResetKernel (initialise BVH and leaf nodes) & 24.62\\
+ bvhConstruct (update parent nodes, create binary radix tree) & 24.94\\
\hline
\end{tabular}
\end{center}
\end{table}

\vspace{-3mm}\subsection{The bvhIntersect kernel}
Putting everything together, the {\small\texttt{bvhFindCollisions}} device code is launched through the {\small\texttt{bvhIntersectKernel}}. The {\small\texttt{bvhRoot}} represents a shared attribute that is initialised by the first thread (within a thread-block) with barrier synchronisation. Akin to version 1 of our implementation, each thread ($i$) iterates over the rays with a stride of $i+\text{gridDim}\times B$ where $B=1024$ is the typical size of a thread-block, and $0\le \text{gridDim} < N_\text{grid}$.

\begin{lstlisting}[frame=single,emph={},emphstyle=\textbf,basicstyle={\ttfamily\footnotesize},escapechar=@]
__global__ void @\textbf{\color{ruby}bvhIntersectionKernel}@(const float* __restrict__ vertices,
                                      const int* __restrict__ triangles,
                                      const float* __restrict__ rayFrom,
                                      const float* __restrict__ rayTo,
                                      const BVHNode* __restrict__ internalNodes,
                                      const AABB* __restrict__ rayBox,
                                      CollisionList* __restrict__ raytriBoxHitIDs,
                                      int* __restrict__ detected,
                                      int numTriangles, int numRays)
{
    //load BVH root node into shared memory
    __shared__ NodePtr bvhRoot;
    __shared__ int stride;
    if (threadIdx.x == 0) {
        bvhRoot = internalNodes[numTriangles-1].childLeft;
        stride = gridDim.x * blockDim.x;
    }
    __syncthreads();

    int threadStartIdx = blockIdx.x * blockDim.x + threadIdx.x;
    int bufferIdx = threadStartIdx;   @\circled{$\star$}@
    //iterate if numRays exceeds dimension of thread-block
    for (int idx = threadStartIdx; idx < numRays; idx += stride) {
        if (idx < numRays) {
            //access thread-specific collision array
            CollisionList &collisions = raytriBoxHitIDs[bufferIdx];
            bvhFindCollisions(vertices, triangles, rayFrom, rayTo, rayBox,
                              bvhRoot, collisions, detected, idx);
        }
    }
}
\end{lstlisting}

Two grid-size configurations, $N_\text{grid}\in\{1,16\}$, were considered. This requires device memory allocation as follows.
\begin{lstlisting}[frame=single,emph={},emphstyle=\textbf,basicstyle={\ttfamily\footnotesize},escapechar=@]
CollisionList *d_hitIDs;
cudaMalloc(&d_hitIDs, nGrids * threadBlockSize * sizeof(CollisionList));
\end{lstlisting}
Essentially, it comes down to a memory and speed trade-off. As the timing measurements show, $N_\text{grid}=1$, results in a 4.7$\times$ improvement relative to version v1. However, with $N_\text{grid}=16$, it results in a 51$\times$ improvement relative to v1---this emphasizes the importance of incorporating an accelerating (tree) structure. It exploits the fact that multiple blocks can be run concurrently on a streaming multiprocessor. For each thread within a given grid-block, the {\small\texttt{bufferIdx}} in line \circled{$\star$} is key to stepping correctly into the dedicated portion of device memory that registers collision. Note: the {\small\texttt{threadStartIdx}} incorporates the {\small\texttt{stride}}, a grid-block specific offset into the rays.

\section{Indicative Results}
\subsection{Processing time}
\begin{table}[h!]
\begin{center}
\renewcommand{\arraystretch}{1.1}
\caption{Timing comparison for the final implementation}\label{tab:compare2}
\begin{tabular}{|l|c|c|c|c|}\hline
& Time (s) & Speed (Mi/s) & GPU\,/\,CPU gain & Change (vs $\ddag$)\\ \hline
\multicolumn{5}{|l|}{\textbf{CPU-based solutions}}\\ \hline
PyVista ({\small\texttt{multi\_ray\_trace}}) with OpenMP & 60.6 & 4832.5$\star$ & $1\times$ & --\\
PyVista ({\small\texttt{ray\_trace}}) & 384.5 & 761.5 & $0.15\times$ & --\\
VTK ({\small\texttt{obbTree.IntersectWithLine}}) & 92.3 & 3171.3 & $0.656\times$ & --\\ \hline
\multicolumn{5}{|l|}{\textbf{Our GPU CUDA implementation}}\\ \hline
Jimenez-like strategy (v1) & 23.729 & 12340.5 & $2.5\times$ & $\ddag$ \\
- with BVH, $N_\text{grid}=1$ (v2.0) & 5.022\ignore{5022ms} & \textbf{58978.1} & $12\times$ & $4.7\times$\\
- with BVH, $N_\text{grid}=16$ (v2.1) preferred & \textbf{\color{ruby}0.463}\ignore{463.282ms} & \textbf{\color{ruby}632071.5} & $130\times$ & $51\times$\\
- with BVH, sorted rays and $N_\text{grid}=16$ (v2.2) & \textbf{0.228}\ignore{228.003ms} & 1284357.7 & $265\times$ & $104\times$\\ \hline
\end{tabular}
\end{center}
\end{table}

In terms of speed, the adopted implementation (v2.1) is $130\times$ faster on a GeForce GTX Titan X GPU than the best CPU-based solution which uses PyVista's {\small\texttt{multi\_ray\_trace}} method. The difference is stark, it completes in 463ms compared with >60s. Furthermore, the rays used for our testing have been randomised. If the rays were first sorted using Morton code, execution divergence can be minimised and the compute time can be halved to about 228ms. However, there is a substantial cost associated with sorting 10M rays. So, one feasible use case would be to sort these rays once and evaluate against multiple surfaces if the line segments remain fixed throughout an experiment.

\subsection{Memory use}
The figures on page~\pageref{list:memuse} show memory usage on the GeForce GTX Titan X GPU device when the program is run. Apart from minor formatting, this output was produced by {\small\texttt{{\color{gray}/usr/local/cuda/bin/}nvprof -\,\!-print-gpu-trace ./gpu\_ray\_surface\_intersect}}. The main observation is that {\small\texttt{bvhIntersectionKernel}} uses 40 registers per thread (a total of 40960 per block) which is less than the 65536 registers available (see Appendix~\ref{sec:gpu-spec}). This kernel uses 12 bytes of static shared memory per block which is tiny compared to the 49152 bytes available. In terms of device global memory, it uses 267.8 MB in the test case which is about 2\% of the 12GB available.

\newpage
\begin{lstlisting}[frame=single,emph={},emphstyle=\textbf,basicstyle={\footnotesize},escapechar=@]
Operation          Object/FunctionName
    Start  Duration  Grid  BlockSz  Regs SSMem DSMem     Size  Throughput  Src/DstMem
========================================================================================
[CUDA memset]      d_crossingDetected
  238.94ms  153.35us     -        -     -     -    -  38.147MB  242.93GB/s  Dev/-
[CUDA memcpy HtoD] d_vertices
  239.13ms  17.088us     -        -     -     -    -  175.38KB  9.7880GB/s  Pg/Dev
[CUDA memcpy HtoD] d_triangles
  239.23ms  30.497us     -        -     -     -    -  343.17KB  10.731GB/s  Pg/Dev
[CUDA memcpy HtoD] d_rayFrom
  239.50ms  23.796ms     -        -     -     -    -  114.44MB  4.6965GB/s  Pg/Dev
[CUDA memcpy HtoD] d_rayTo
  263.44ms  22.907ms     -        -     -     -    -  114.44MB  4.8788GB/s  Pg/Dev
[kernel code]      lib_rsi::rbxKernel
  286.39ms  4.2089ms (9766,) (1024,)   17    0B   0B         -           -   -/-
[CUDA memcpy HtoD] d_morton
  310.83ms  21.793us     -        -     -     -    -  228.78KB  10.012GB/s  Pg/Dev
[CUDA memcpy HtoD] d_sortedTriangleIDs
  310.87ms  11.328us     -        -     -     -    -  114.39KB  9.6302GB/s  Pg/Dev
[kernel code]      lib_bvh::bvhResetKernel
  310.89ms  83.682us   (29,) (1024,)   26    0B   0B         -           -   -/-
[kernel code]      bvhConstructKernel
  310.98ms  207.94us   (29,) (1024,)   26    0B   0B         -           -   -/-
[kernel code]      @\textbf{lib\_bvh::bvhIntersectionKernel}@
  311.20ms  489.58ms   (16,) (1024,)   40   12B   0B         -           -   -/-
[CUDA memcpy DtoH] h_crossingDetected
  800.81ms  3.2842ms     -        -     -     -    -  38.147MB  11.343GB/s Dev/Pg

Regs:  Number of registers used per CUDA thread. This number includes registers used
       internally by the CUDA driver/tools and can be more than what the compiler shows
SSMem: Static shared memory allocated per CUDA block.
DSMem: Dynamic shared memory allocated per CUDA block.
Src/DestMem: The type of source/destination memory accessed by memory operation/copy
Abbreviations:  Pg=Pageable, Dev=Device
\end{lstlisting}\label{list:memuse}

\newpage\appendix
\section{Appendix: System\,/\,device properties}
\subsection{GPU specification}\label{sec:gpu-spec}
\begin{lstlisting}[frame=single,emph={},emphstyle=\textbf,basicstyle={\ttfamily\footnotesize},escapechar=@]
CUDA Device Query (Runtime API) version (CUDART static linking)

Detected 1 CUDA Capable device(s)

Device 0: @\textbf{"GeForce GTX TITAN X"}@
  CUDA Driver Version / Runtime Version          11.2 / 11.2
  CUDA Capability Major/Minor version number:    5.2
  Total amount of global memory:                 12213 MBytes (12806062080 bytes)
  (24) Multiprocessors, (128) CUDA Cores/MP:     3072 CUDA Cores
  GPU Max Clock rate:                            1076 MHz (1.08 GHz)
  Memory Clock rate:                             3505 Mhz
  Memory Bus Width:                              384-bit
  L2 Cache Size:                                 3145728 bytes
  Maximum Texture Dimension Size (x,y,z)         1D=(65536), 2D=(65536, 65536),
                                                 3D=(4096, 4096, 4096)
  Maximum Layered 1D Texture Size, (num) layers  1D=(16384), 2048 layers
  Maximum Layered 2D Texture Size, (num) layers  2D=(16384, 16384), 2048 layers
  Total amount of constant memory:               65536 bytes
  Total amount of shared memory per block:       49152 bytes
  Total number of registers available per block: 65536
  Warp size:                                     32
  Maximum number of threads per multiprocessor:  2048
  Maximum number of threads per block:           1024
  Max dimension size of a thread block (x,y,z): (1024, 1024, 64)
  Max dimension size of a grid size    (x,y,z): (2147483647, 65535, 65535)
  Maximum memory pitch:                          2147483647 bytes
  Texture alignment:                             512 bytes
  Concurrent copy and kernel execution:          Yes with 2 copy engine(s)
  Run time limit on kernels:                     No
  Integrated GPU sharing Host Memory:            No
  Support host page-locked memory mapping:       Yes
  Alignment requirement for Surfaces:            Yes
  Device has ECC support:                        Disabled
  Device supports Unified Addressing (UVA):      Yes
  Device PCI Domain ID / Bus ID / location ID:   0 / 2 / 0
  Compute Mode:
     @\text{< Default (multiple host threads can use ::cudaSetDevice() with device simultaneously) >}@

deviceQuery, CUDA Driver = CUDART, CUDA Driver Version = 11.2, 
CUDA Runtime Version = 11.2, NumDevs = 1, Device0 = GeForce GTX TITAN X
Result = PASS

NVIDIA-SMI driver version is 460.27.04.
\end{lstlisting}

\subsection{Host machine}
Linux OS (RedHat 7.9 x86\_64) with 16x Intel\textsuperscript{\textregistered} Xeon\textsuperscript{\textregistered} Platinum 8259CL CPU @ 2.50GHz. Experiments with PyVista and VTK were run on a Windows machine with Intel\textsuperscript{\textregistered} Core\textsuperscript{TM}  i7-8665U CPU @ 1.90GHz, 4 cores/8 logical processors and 32 GB RAM.

\section{Source code}
The open-source code is distributed under the BSD 3-clause license and is available at \url{https://github.com/raymondleung8/gpu-ray-surface-intersection-in-cuda} \cite{leung2022gpursi}.

\section*{Acknowledgements}
This work is supported by the Australian Centre for Field Robotics and the Rio Tinto Centre for Mine Automation.

\bibliographystyle{unsrt}
\bibliography{references.bib}

\end{document}